\documentclass[aps,prd,floatfix,12pt]{revtex4}
\usepackage{epsfig}
\usepackage{bm}
\usepackage{dcolumn}
\usepackage{latexsym}

\begin{document}
   
\preprint{\rightline{ANL-HEP-PR-10-xx}}
   
\title{Thermodynamics of lattice QCD with 2 flavours of colour-sextet quarks:
  A model of walking/conformal Technicolor}
   
\author{J.~B.~Kogut}
\affiliation{Department of Energy, Division of High Energy Physics, Washington,
DC 20585, USA}
 \author{\vspace{-0.2in}{\it and}}
\affiliation{Dept. of Physics -- TQHN, Univ. of Maryland, 82 Regents Dr.,
College Park, MD 20742, USA}
\author{D.~K.~Sinclair}
\affiliation{HEP Division, Argonne National Laboratory, 9700 South Cass Avenue,
Argonne, IL 60439, USA}

\begin{abstract}
QCD with two flavours of massless colour-sextet quarks is considered as a model
for conformal/walking Technicolor. If this theory possess an infrared fixed 
point, as indicated by 2-loop perturbation theory, it is a conformal(unparticle)
field theory. If, on the other hand, a chiral condensate forms on the 
weak-coupling side of this would-be fixed point, the theory remains confining.
The only difference between such a theory and regular QCD is that there is a
range of momentum scales over which the coupling constant runs very slowly
(walks). In this first analysis, we simulate the lattice version of QCD with
two flavours of staggered quarks at finite temperatures on lattices of 
temporal extent $N_t=4$ and $6$. The deconfinement and chiral-symmetry
restoration couplings give us a measure of the scales associated with
confinement and chiral-symmetry breaking. We find that, in contrast to what is
seen with fundamental quarks, these transition couplings are very different.
$\beta=6/g^2$ for each of these transitions increases significantly from
$N_t=4$ and $N_t=6$ as expected for the finite temperature transitions of an
asymptotically-free theory. This suggests a walking rather than a conformal
behaviour, in contrast to what is observed with Wilson quarks. In contrast to
what is found for fundamental quarks, the deconfined phase exhibits states in
which the Polyakov loop is oriented in the directions of all three cube roots
of unity. At very weak coupling the states with complex Polyakov loops undergo
a transition to a state with a real, negative Polyakov loop.

\end{abstract}

\maketitle

\section{Introduction}

The advent of the LHC era has led to renewed interest in extensions of the
Standard Model which have a strongly-coupled symmetry-breaking (Higgs) sector.
The most promising of these theories are the so-called Technicolor theories
\cite{Weinberg:1979bn,Susskind:1978ms}. These are QCD-like theories with
massless fermions (techni-quarks) in which the Goldstone bosons of
spontaneously-broken chiral symmetry (techi-pions) play the r\^{o}le of the
Higgs field, giving masses to the $W$ and $Z$. Such models need to be extended
in order to also give masses to the quarks and leptons. Phenomenological
difficulties with such models, such as flavour-changing neutral currents which
are too large, can be avoided if the fermion content of a candidate (extended)
Technicolor theory is such that the running gauge coupling constant evolves
very slowly. Such theories are referred to as Walking Technicolor models
\cite{Holdom:1981rm,Yamawaki:1985zg,Akiba:1985rr,Appelquist:1986an}.

Let us consider a Yang-Mills gauge theory with $N_f$ fermions in a specified
(not-too-large) representation of the ``colour'' group. The evolution of the
coupling constant $g$ in such a theory is described by the Callan-Symanzik
beta function $\beta(g)$ defined by
\begin{equation}
\beta(g) = \mu{\partial g \over \partial \mu} =
- \beta_0{g^3 \over (4\pi)^2} - \beta_1{g^5 \over (4\pi)^4}...
\end{equation}
where $\mu$ is the momentum scale at which the running coupling constant
$g(\mu)$ is defined. $\beta_0$,$\beta_1$,... are given by perturbation theory.
For $N_f$ sufficiently small, $\beta_0$ and $\beta_1$ are both positive, 
the theory is asymptotically free and confining, and chiral symmetry is 
spontaneously broken. There exists some value of $N_f$ above which $\beta_0$
(and $\beta_1$) are negative and asymptotic freedom is lost. Between these
two regimes is a range of $N_f$ over which $\beta_0$ is positive and $\beta_1$
is negative. In this range the theory remains asymptotically free, but if
this 2-loop beta function describes the physics, it has a second zero which
is an infrared (IR) fixed point and the theory is a conformal (unparticle)
field theory. There is, however, another possibility. If the coupling becomes
strong enough that a chiral condensate forms before this would-be IR fixed 
point can be reached, the fermions will become less effective at screening
technicolor. $\beta$ will then start to decrease again, and the theory will be
confining in the infrared. There will, however, be a range of $\mu$ over which
$\beta$ is small and $g$ evolves only slowly, i.e. it walks. Since the 
formation of a chiral condensate which spontaneously breaks chiral symmetry is
a non-perturbative process, the boundary between conformal and walking 
behaviour cannot be determined perturbatively. Lattice gauge theory simulations
can enable one to decide between these two options for a theory with a 
specified gauge group, fermion technicolor representation and $N_f$. 

For $SU(N)$ gauge theories, the most promising candidates for walking behaviour
have fermions in the fundamental, adjoint, symmetric 2-index tensor or
antisymmetric 2-index tensor representations of the gauge group. A good summary
of what is known is given in reference \cite{Dietrich:2006cm}. Estimates of
the value of $N_f$ below which a gauge theory walks and above which it is
conformal have been made using various methods , none of which can be
guaranteed to capture the full non-perturbative behaviour of QCD-like theories
\cite{Appelquist:1988yc,Sannino:2004qp,Poppitz:2009uq,Armoni:2009jn,
Ryttov:2007cx,Antipin:2009wr}. This has led people to use lattice gauge theory
simulations to study this boundary. There have been a number of simulations of
QCD with $N_f$ fundamental quarks, with $N_f$ large enough that conformal or
walking behaviour might be expected
\cite{Kogut:1985pp,Fukugita:1987mb,Ohta:1991zi,Kim:1992pk,Brown:1992fz,
Iwasaki:1991mr,Iwasaki:2003de,Deuzeman:2008sc,Deuzeman:2009mh,
Appelquist:2009ty,Appelquist:2007hu,Jin:2008rc,Jin:2009mc,Fodor:2009wk,
Fodor:2009ff,Yamada:2009nt}. While progress has been made, the boundary value
for $N_f$ is still uncertain. Some simulations have been made of $SU(2)$ gauge
theory with 2 Dirac flavours of adjoint fermions
\cite{Catterall:2007yx,Catterall:2008qk,Catterall:2009sb,DelDebbio:2008zf,
DelDebbio:2009fd,Bursa:2009we,Hietanen:2008mr,Hietanen:2009az}. While early
indications are that this is a conformal field theory, there is still a great
deal of uncertainty. 

For QCD with colour-sextet (symmetric tensor) quarks, $\beta_1$ changes sign
at $N_f=1\frac{28}{125}$, and asymptotic freedom is lost at
$N_f=3\frac{3}{10}$. Hence only $N_f=2$ and $N_f=3$ lie in the domain of
interest. $N_f=3$ is just below the value where asymptotic freedom is lost and
is thus expected to be conformal. This leaves $N_f=2$ as the most likely
candidate for walking behaviour. DeGrand, Shamir and Svetitsky have simulated
lattice QCD with 2 flavours of colour-sextet Wilson quarks
\cite{Shamir:2008pb,DeGrand:2008kx,DeGrand:2009hu}. Their initial results
suggest that this is a conformal field theory. Fodor et al. have performed
some preliminary simulations of lattice QCD with 2 sextet quarks using
domain-wall quarks \cite{Fodor:2008hm}.

We are performing simulations of lattice QCD with 2 colour-sextet staggered
quarks. Staggered quarks have the advantage over Wilson quarks in having a
simple chiral order parameter, and no parameter tuning is needed to find the
chiral limit. We are performing simulations at finite temperatures, where the
deconfinement and chiral-symmetry restoration temperatures can be used as
measures of the scales of confinement and chiral-symmetry breaking
respectively. By varying the number of time slices $N_t$ we can determine
whether these transitions scale as finite temperature transitions or whether
they are bulk transitions. Preliminary results of these simulations were
presented at Lattice 2009 \cite{Sinclair:2009ec}.

Our simulations indicate that, whereas for fundamental quarks the deconfinement
and chiral-symmetry restoration transitions appear coincident, for
colour-sextet quarks these transitions are well separated. Chiral-symmetry
restoration occurs at a much weaker coupling than deconfinement. This differs
from what is seen with Wilson quarks by DeGrand, Shamir and Svetitsky
\cite{DeGrand:2008kx} where the two transitions appear coincident. Such a
separation has been reported in earlier simulations with colour-adjoint quarks
\cite{Karsch:1998qj,Engels:2005te}, and in $SU(2)$ gauge theory with 
colour-adjoint quarks \cite{Kogut:1985xa}. Both transitions move to
significantly weaker couplings when $N_t$ is increased from $4$ to $6$, which
is what would be expected for finite temperature transitions governed by
asymptotic freedom. This in turn favours the walking scenario.

In the deconfined region, just above the deconfinement transition, we find 3
states, where the Wilson Line (Polyakov Loop) is oriented in the directions of
the 3 cube roots of unity, similar to what occurs for quenched QCD or QCD with
adjoint quarks. For $N_t=4$ only the state with a real positive Polyakov Loop
appears stable. The other two states, while long-lived, appear to be only
metastable. For $N_t=6$, all 3 states appear to be stable. Between the
deconfinement and chiral-symmetry restoration transitions there is another
transition where the 2 states with complex Polyakov Loops disorder into a
state with a real, negative Polyakov Loop. Machtey and Svetitsky have argued
that such additional states where the Polyakov Loop has arguments $\pm 2\pi/3$
and $\pi$ are to be expected, and have presented evidence for their existence
and metastability in simulations using Wilson quarks \cite{Machtey:2009wu}.

In section~2 we discuss our simulations. Our results are described in section~3.
Section~4 presents discussions and conclusions and identifies directions for
ongoing and future investigation.

\section{Lattice simulations with sextet quarks}

For our simulations we use the simple Wilson gauge action
\begin{equation}
S_g=\beta \sum_\Box \left[1-\frac{1}{3}{\rm Re}({\rm Tr}UUUU)\right].
\end{equation}
The fermion action is based on the unimproved staggered-quark action written
formally as
\begin{equation}
S_f=\sum_{sites}\left[\sum_{f=1}^{N_f/4}\psi_f^\dagger[D\!\!\!\!/+m]\psi_f
\right]
\end{equation}
where $D\!\!\!\!/ = \sum_\mu \eta_\mu D_\mu$ with 
\begin{equation}
D_\mu \psi(x) = \frac{1}{2}[U^{(6)}_\mu(x)\psi(x+\hat{\mu})-
                            U^{(6)\dagger}_\mu(x-\hat{\mu})\psi(x-\hat{\mu})].
\end{equation}
To allow tuning the number of flavours to values of $N_f$ which are not 
multiples of 4, and to use a positive-definite operator for the transition to
pseudofermions, this is replaced with
\begin{equation}
S_f=\sum_{sites}\chi^\dagger\{[D\!\!\!\!/+m][-D\!\!\!\!/+m]\}^{N_f/8}\chi.
\end{equation}
We use the RHMC algorithm for our simulations in which the fractional powers
of the positive-definite Dirac operator are approximated to any desired accuracy
by a rational approximation, and each trajectory is subjected to a global
Metropolis accept/reject step, thus removing all dependence on the updating
increment.

     We perform simulations on $8^3 \times 4$, $12^3 \times 4$ and 
$12^3 \times 6$ lattices, over a range of values of $\beta=6/g^2$ which covers
all transitions. For each lattice size we perform runs at quark mass $m=0.005$,
$m=0.01$ and $m=0.02$ in lattice units, to allow us to access the chiral limit.
Away from the transitions our run lengths are typically 10,000 length-1 
trajectories for a given $m$ and $\beta$. Near deconfinement transitions run 
lengths of 50,000 to 200,000 trajectories are used for each $\beta$ and $m$. 
Some runs of 50,000 trajectories have also been used near the transitions from
negative Polyakov Loop states to complex Polyakov Loop states.

     We create deconfined states with positive Polyakov Loops by starting
a run at $\beta=7.0$ (weak coupling) from a completely ordered state (all
$U$s equal to the identity matrix). The configurations from these runs are
used to start runs at progressively smaller $\beta$s (and masses). The states
with negative Polyakov Loops are obtained by starting a run at $\beta=7.0$
with all $U$s equal to the identity matrix, except for the timelike $U$s on
a single time slice, which are set to the matrix ${\rm diag}(1,-1,-1)$.

     The triplet Wilson Line (Polyakov Loop) is used to identify the position
of the deconfinement transition. The chiral-symmetry restoring phase transition
occurs at that $\beta$ above which the chiral condensate 
($\langle\bar{\psi}\psi\rangle$) vanishes in the chiral ($m \rightarrow 0$)
limit. Since the chiral extrapolation is difficult to perform with the masses
we use, we estimate the position of the chiral transition from the positions
of the peaks in the chiral susceptibilities $\chi_{\bar{\psi}\psi}$ as 
functions of mass.
\begin{equation}
\chi_{\bar{\psi}\psi} = V\left[\langle(\bar{\psi}\psi)^2\rangle
                      -        \langle\bar{\psi}\psi\rangle^2\right]
\label{eqn:chi}
\end{equation}
where the $\bar{\psi}\psi$s in this formula are lattice averaged quantities
and $V$ is the space-time volume of the lattice. Since we use stochastic
estimators for $\bar{\psi}\psi$, we obtain an unbiased estimator for this
quantity by using several independent estimates for each configuration (5, in
fact). Our estimate of $(\bar{\psi}\psi)^2$ is then given by the average of
the (10) estimates which are `off diagonal' in the noise.

\section{Results from simulations}

\subsection{$N_t=4$}

\begin{figure}[hb]
\epsfxsize=6in
\epsffile{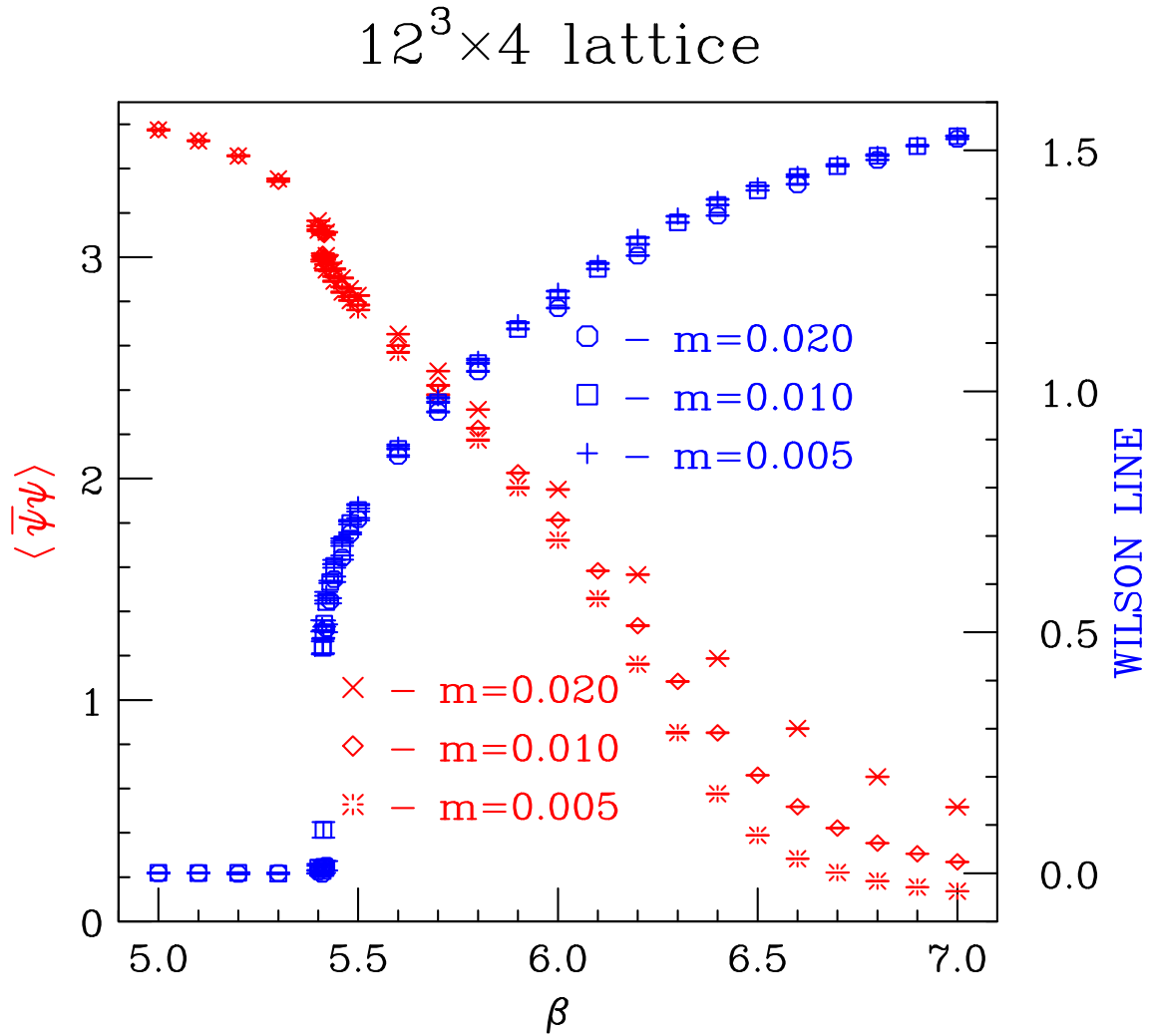}
\caption{Wilson line (Polyakov Loop) and $\langle\bar{\psi}\psi\rangle$
as functions of $\beta$ on a $12^3 \times 4$ lattice.}
\label{fig:wil-psi_12x4}
\end{figure}

We perform simulations at a selection of $\beta$ values in the range
$5.0 \le \beta \le 7.0$ on $8^3 \times 4$ and $12^3 \times 4$ lattices. 
For each of the 3 chosen masses ($0.005$, $0.01$, $0.02$) the values of the
Wilson Line and chiral condensate on the $8^3 \times 4$ and $12^3 \times 4$
lattices are so close that we conclude that finite size effects are negligible.
Figure~\ref{fig:wil-psi_12x4} shows the Wilson Line and chiral condensates
as functions of $\beta$ for each of the 3 masses ($0.005$, $0.01$, $0.02$) on
a $12^3 \times 4$ lattice. In the deconfined phase, we have included only
those states where the Wilson Line is real and positive, noting that runs
which start in a state with a positive Wilson Line remain in this state for
the duration of the run. We have not included the results for the $8^3 \times
4$ lattice, since at the resolution of figure~\ref{fig:wil-psi_12x4}
the points for the 2 lattice sizes would be virtually indistinguishable.

It is clear that the Wilson Line is very small below $\beta \approx 5.4$,
and rises rapidly shortly after this value for all 3 quark masses. This is
taken as a signal for the deconfinement transition. It is also clear that
chiral symmetry remains broken well beyond this point. Thus the deconfinement
and chiral-symmetry restoration transitions are far apart, unlike what is
observed for Wilson quarks, where they appear to be coincident 
\cite{DeGrand:2008kx}.

\begin{figure}[htb]
\epsfxsize=5.0in                  
\epsffile{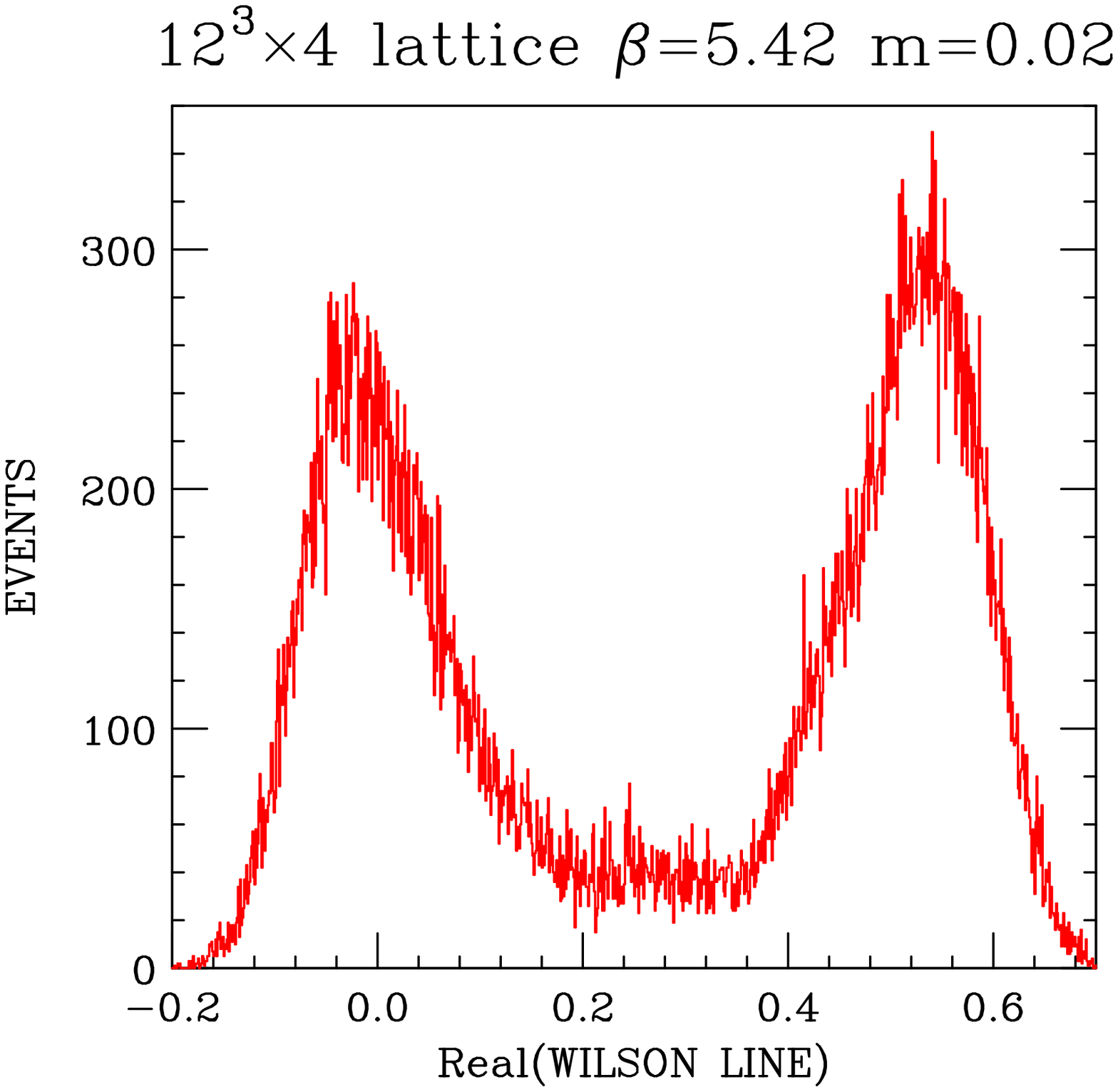}
\caption{Histogram of the Wilson Line at $\beta=5.42$, $m=0.02$ on a 
$12^3 \times 4$ lattice.}
\label{fig:wilhist-5.42}
\end{figure}

      Figure~\ref{fig:wilhist-5.42} shows a histogram of the Wilson Line at
$\beta=5.42$, $m=0.02$, showing a clear two-state signal, suggesting that
this $\beta$ is very close to the transition. The separation of the two states 
suggests that this transition is a first-order phase transition. We conclude
that at $m=0.02$ the deconfinement transition is at $\beta=\beta_d=5.420(5)$.
For $m=0.01$, two-state signals are seen at $\beta=5.411$ and $\beta=5.412$
leading to an estimate $\beta_d=5.4115(5)$. Finally we note that for $m=0.005$,
$\beta=5.4$ appears to lie below the transition while $\beta=5.41$ appears to  
be above the transition leading to an estimate $\beta_d=5.405(5)$. Thus the
mass dependence of the deconfinement $\beta$, $\beta_d$, is very weak.

\begin{figure}[htb]
\epsfxsize=5.0in
\epsffile{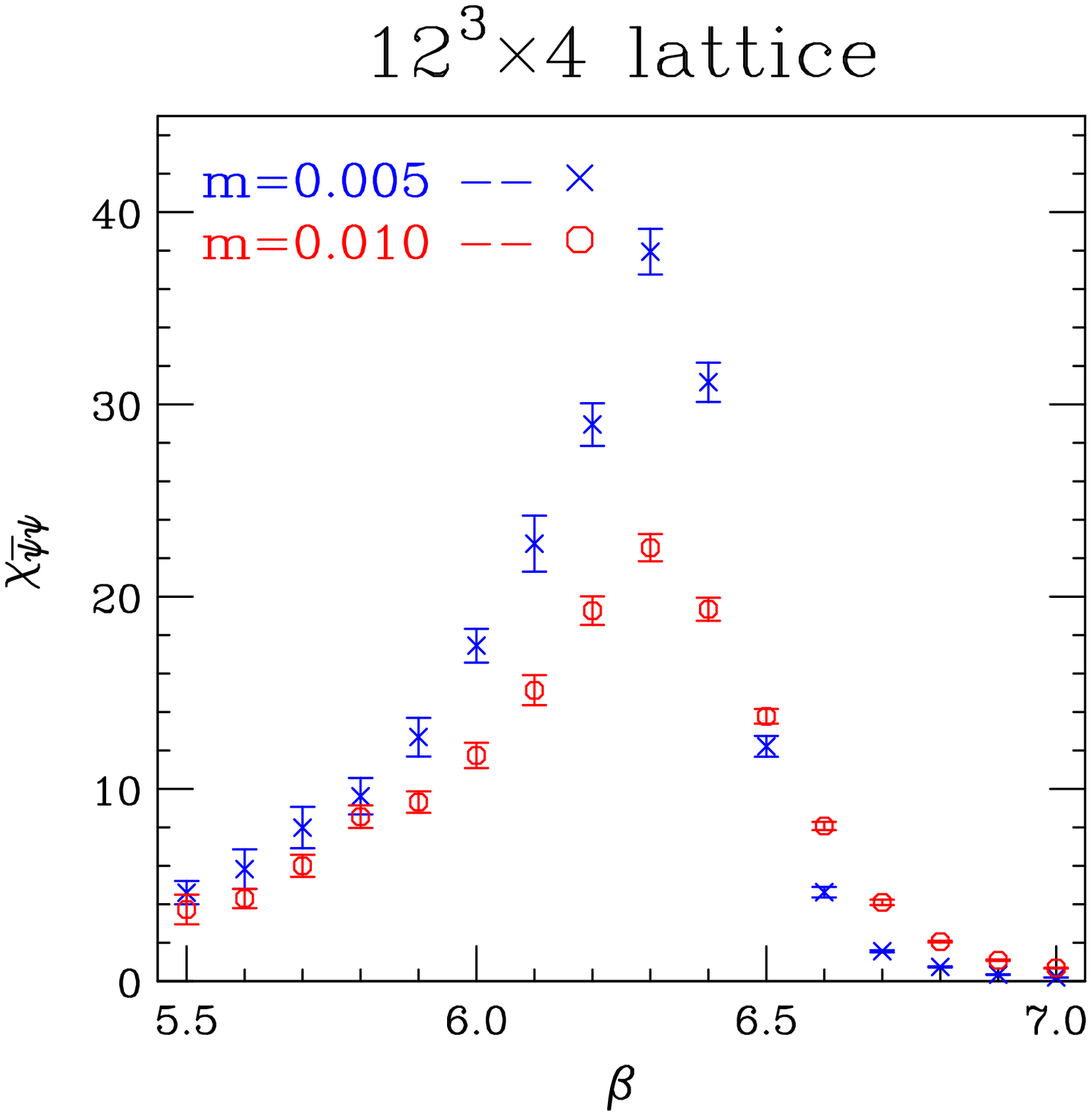}
\caption{Chiral susceptibilities $\chi_{\bar{\psi}\psi}$ as functions of $\beta$
on a $12^3 \times 4$ lattice for $m=0.005,0.01$, for a $\beta$ range which 
includes the chiral transition.}
\label{fig:chi4}
\end{figure}

      Now let us turn to the chiral transition. Because this is only expected
to be a phase transition at $m=0$, and the crossover becomes smoother as the
quark mass is increased, it is difficult to determine its position directly
from the chiral condensate at the masses we use. This is made more difficult
since it is clear from the measured condensates as functions of mass, that
the mass dependence is far from linear, making it extrapolating to $m=0$
difficult if not impossible. We therefore use the peak in the chiral 
susceptibility defined in equation~\ref{eqn:chi} as an estimate of the position
of the crossover for finite mass. This is shown in figure~\ref{fig:chi4} for
the two smallest masses.
From these graphs we estimate that the transition occurs at $\beta=6.30(5)$
for both $\beta$s. We thus estimate that the phase transition at $m=0$ occurs
at $\beta=\beta_\chi=6.3(1)$. Note that the spacing of the $\beta$s in this 
range is too large to allow us to use Ferrenberg-Swendsen reweighting to 
determine the positions of these transitions with more resolution. (The
distributions of values of the plaquette action for adjacent $\beta$s do not
overlap in this region.)

\begin{figure}[htb]
\epsffile{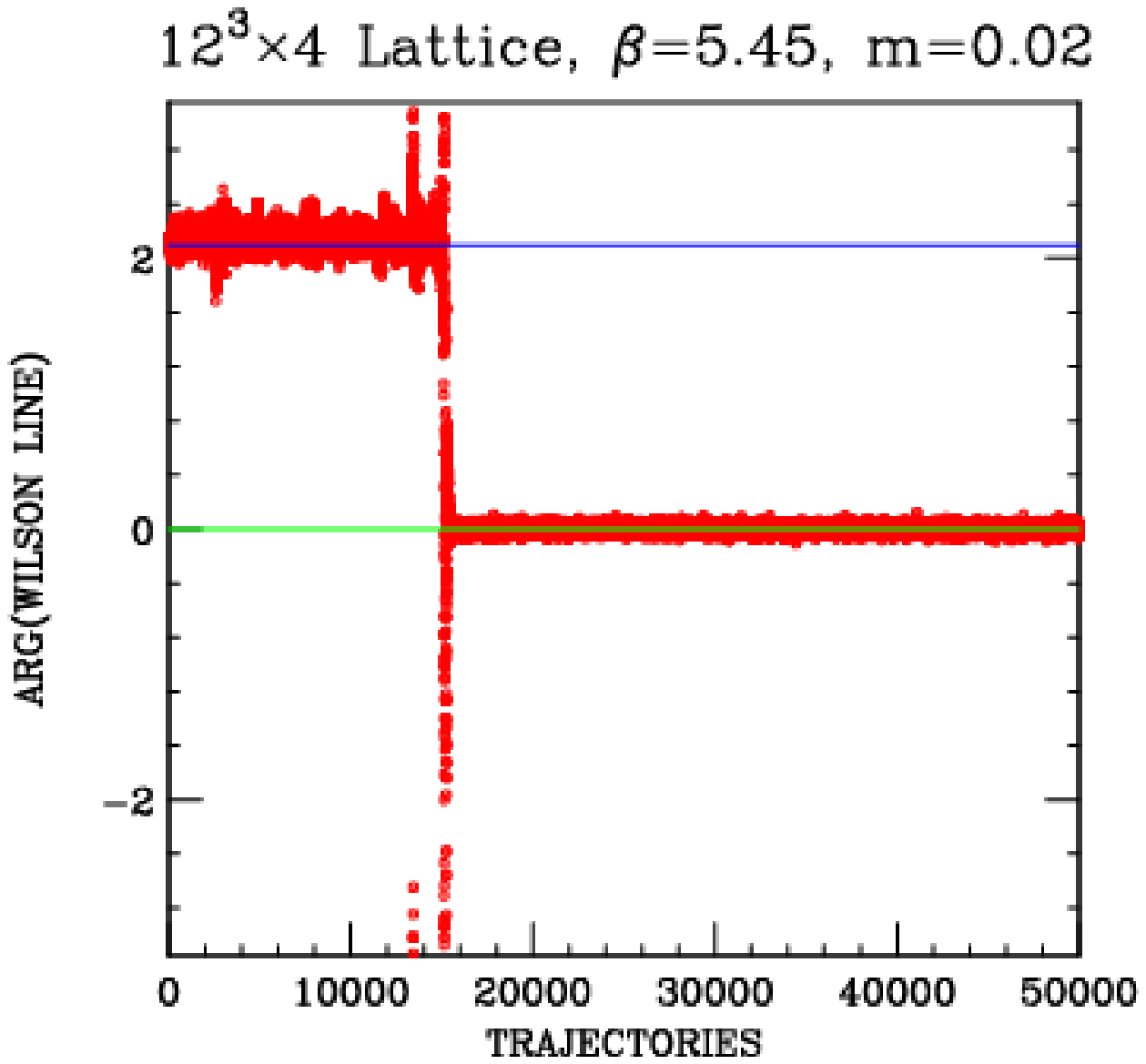}
\caption{Time evolution of the argument of the Wilson Line (Polyakov Loop)
at $m=0.02$, $\beta=5.45$ showing a decay of a state with a complex Wilson
Line (argument close to $2\pi/3$) to a state with a real positive Wilson
Line (argument close to $0$).}
\label{fig:meta}
\end{figure}
At each of our two larger masses, we have performed a series of runs starting
from $\beta=7.0$ with a negative Wilson Line. From $\beta=7.0$ down to 
$\beta=6.0$, the Wilson Lines remain negative over the length of the 10,000
trajectory run for each $(\beta,m)$. By $\beta=5.8$, these states have 
transitioned to states in which the Wilson Line is oriented in the direction of
one of the complex cube roots of unity. Hence we deduce that there is a
transition at $\beta \approx 5.9$. More discussion of this transition is to be
found in the $N_t=6$ subsection. On the $12^3 \times 4$ lattice with $m=0.02$
these states with complex Wilson Lines persist down to $\beta=5.48$, and
appear stable over at least 50,000 trajectories. For $m=0.01$ these persist
down to $\beta=5.46$. For $\beta \le 5.46$ at $m=0.02$ or $\beta \le 5.45$ at
$m=0.01$ but above the deconfinement transition, these states with complex
Wilson Lines appear to be metastable and eventually decay into states with
positive Wilson Lines. Figure~\ref{fig:meta} shows an example of such
metastability in our $12^3 \times 4$ simulations. As is to be expected, this
metastability starts at larger $\beta$ values on an $8^3 \times 4$ lattice. We
have observed no cases where configurations with positive Wilson Lines evolve
to configurations with complex Wilson Lines for $\beta$ values above the
deconfinement transition.

\subsection{$N_t=6$}

We perform simulations on a $12^3 \times 6$ lattice at quark masses $m=0.005$,
$m=0.01$ and $m=0.02$ for values of $\beta=6/g^2$ in the range 
$5.0 \le \beta \le 7.2$. We perform two series of runs starting at $\beta=7.0$,
for $m=0.01,0.02$. The first starts with the Wilson Line real and positive, and
the second with the Wilson Line negative. For the lowest quark mass $m=0.005$
we only perform one set of runs starting with a positive Wilson Line at 
$\beta=7.0$.

\begin{figure}[htb]
\epsffile{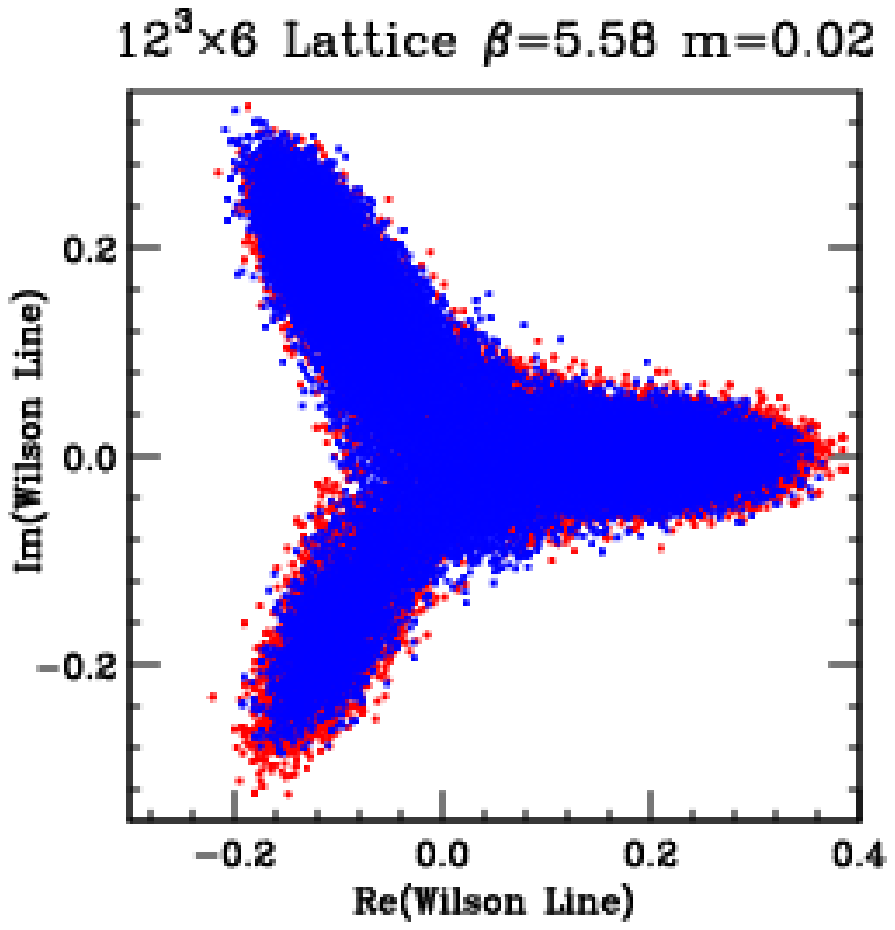}
\epsffile{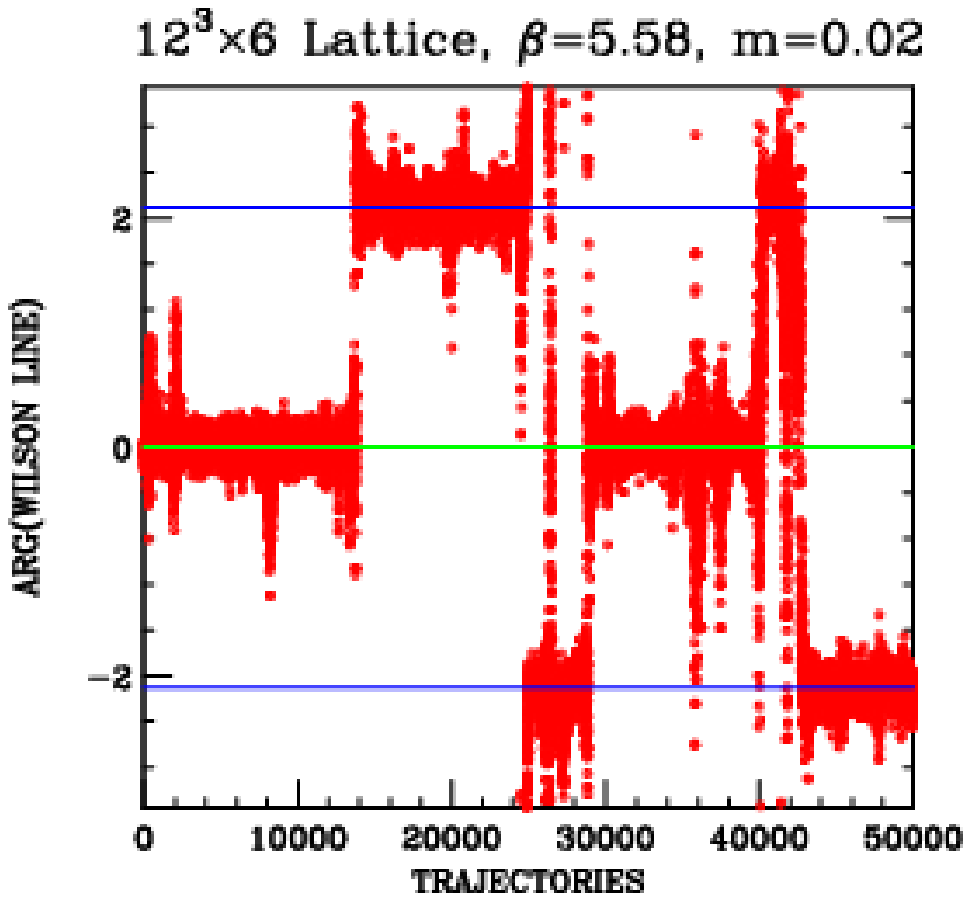}
\caption{a) Scatterplot of Wilson Lines for $m=0.02$ and $\beta=5.58$ on a
$12^3 \times 6$ lattice showing the 3-state signal. \newline 
b) `Time' evolution of the argument(phase) of the Wilson Lines for one of the 
2 runs included in part (a).}
\label{fig:Z3}
\end{figure}
Above the deconfinement transition we again see a 3-state signal where the
Wilson Lines orient themselves in the directions of one of the 3 cube roots of
unity. A scatterplot showing such a 3-state signal is shown in
figure~\ref{fig:Z3}a. Unlike the $N_t=4$ case, there is no sign of
metastability and transitions between all 3 states are seen over the duration
of our runs, up to $\beta$ values far enough above the deconfinement transition
that the mean relaxation time between tunnelings exceeds the lengths of our
runs (50,000 trajectories). Figure~\ref{fig:Z3}b shows an example of such
tunnelings. Thus to make meaningful measurements of the Wilson Line, we need
to separate these 3 states. This we do by binning the Wilson Lines measured at
the end of each trajectory according to their phase $\phi$ into bins 
$-\pi < \phi < -\pi/3$, $-\pi/3 < \phi < \pi/3$, $\pi/3 < \phi < \pi$. To
increase our statistics, we use symmetry to include the complex conjugates of
the Wilson Lines in the first of these bins, with the Wilson Lines in the last
of these bins. Other observables are binned according to the values of the
corresponding Wilson Lines.

\begin{figure}[htb]
\epsfxsize=4.0in
\epsffile{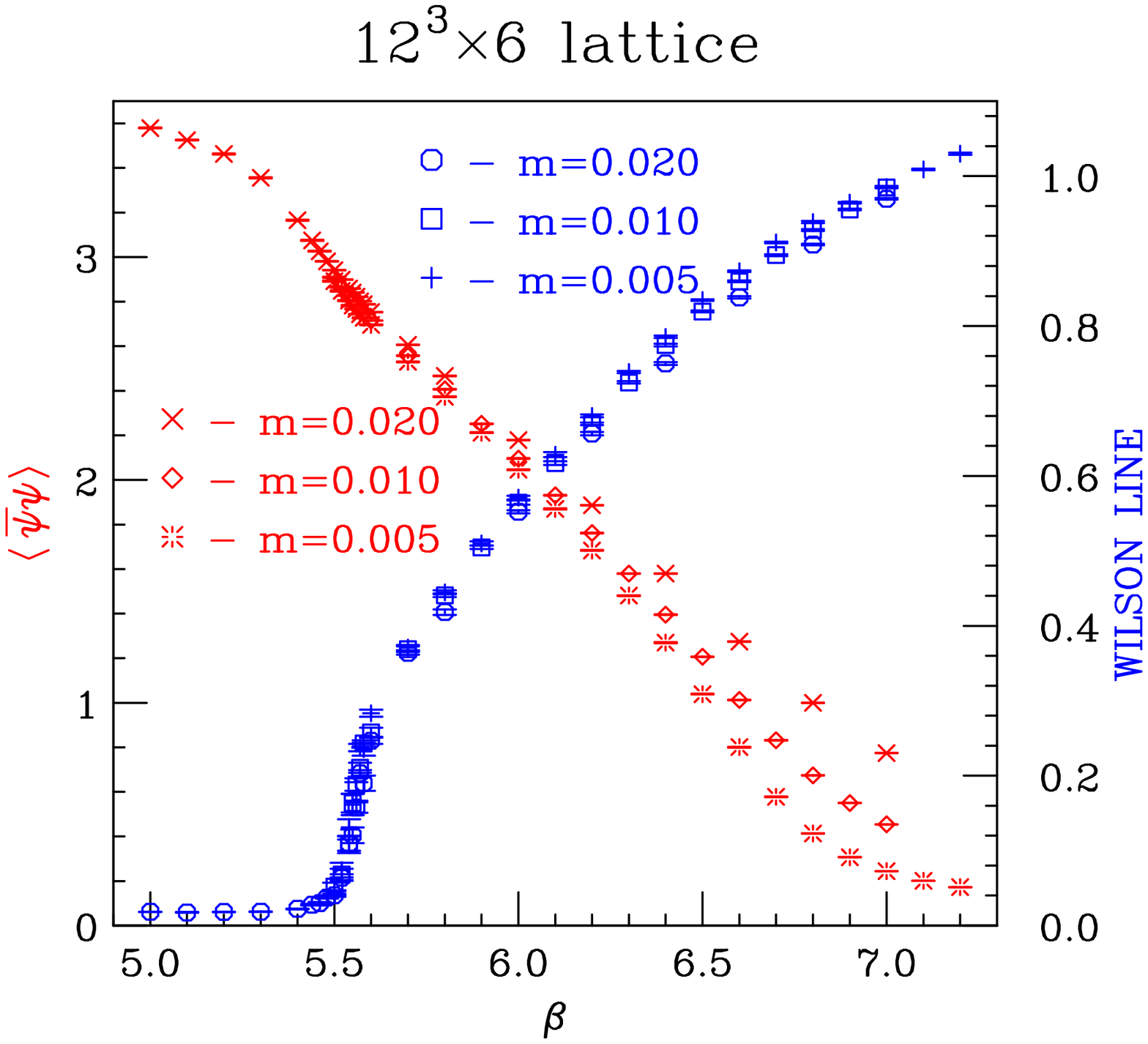}
\epsfxsize=4.0in                                                           
\epsffile{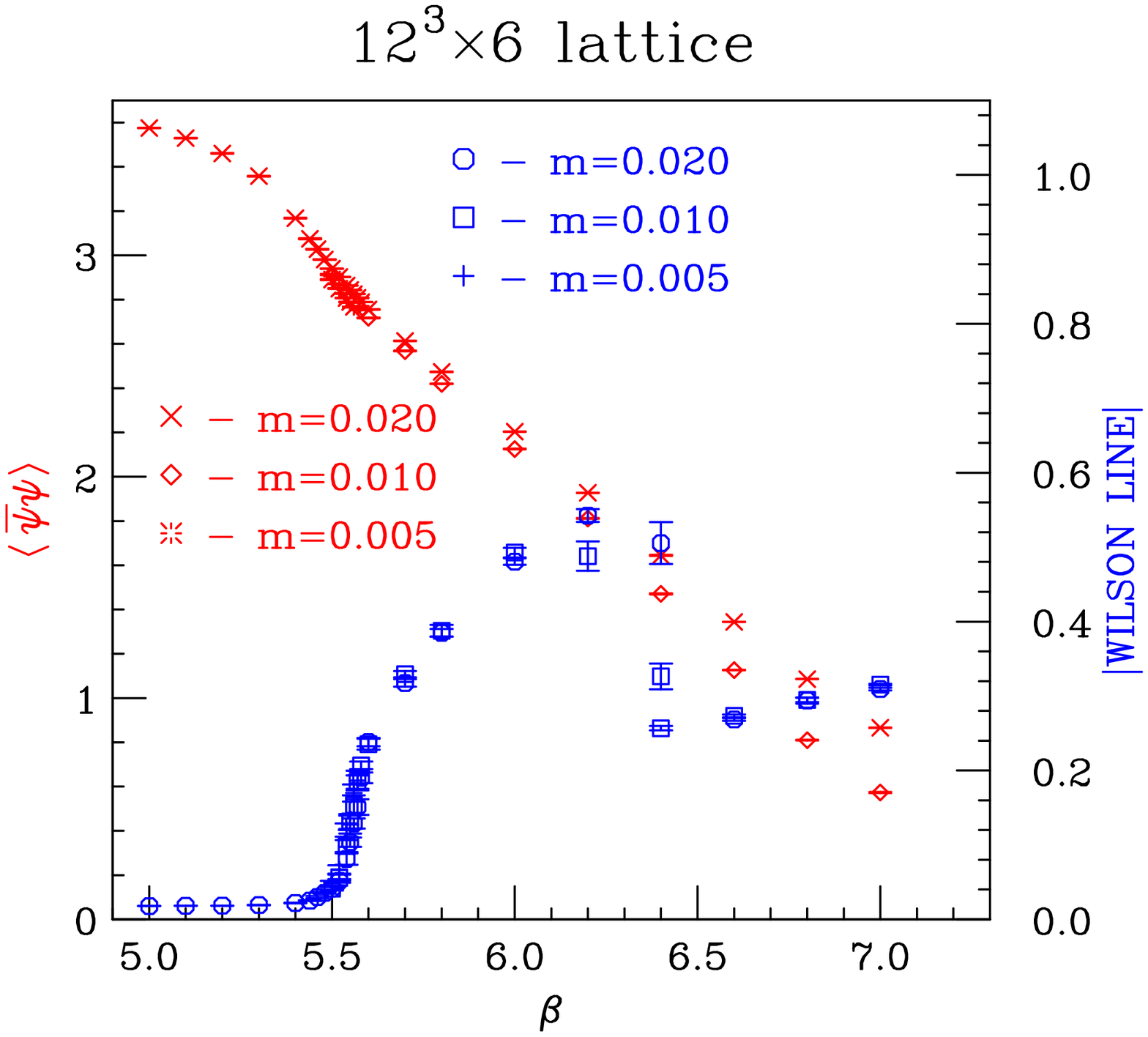}    
\caption{a) Wilson Line and chiral condensate for the state with a real positive
Wilson Line as functions of $\beta$ for each of the 3 masses on a 
$12^3 \times 6$ lattice. \newline
b) Magnitude of the Wilson Line and chiral condensate for the states
with a complex or a real negative Wilson Line as functions of $\beta$ for each
of the 3 masses on a $12^3 \times 6$ lattice.}
\label{fig:wil-psi_4x6}
\end{figure}
Figure~\ref{fig:wil-psi_4x6}a shows the Wilson Lines (Polyakov Loops) and 
chiral condensates as functions of $\beta$ for all 3 masses from our simulations
on a $12^3 \times 6$ lattice for the states with positive Wilson Lines
($-\pi/3 < \phi < \pi/3$). Figure~\ref{fig:wil-psi_4x6}b shows the magnitudes
of the average Wilson Lines and the chiral condensates for the states with
complex or negative Wilson Lines. The deconfinement transition manifests itself
as a rapid increase in the Wilson Line, which is clearly seen at a $\beta$ just
above $5.5$. The fact that the $Z_3$ centre symmetry is broken manifests 
itself in the difference between the magnitudes of the Wilson Lines in the
positive Wilson Line and complex Wilson Line states. The rapid change in the
magnitude of the complex/negative Wilson line between $\beta=6.2$ and 
$\beta=6.6$ marks the transition between a state whose Wilson Line is oriented
in the direction of one of the complex cube roots of unity and one where the
Wilson Line is real and negative. The chiral transition, above which
$\langle\bar{\psi}\psi\rangle$ vanishes in the chiral ($m \rightarrow 0$)
limit, clearly occurs at a $\beta$ appreciably larger $\beta_d$.

\begin{figure}[htb]
\epsfxsize=3.2in
\epsffile{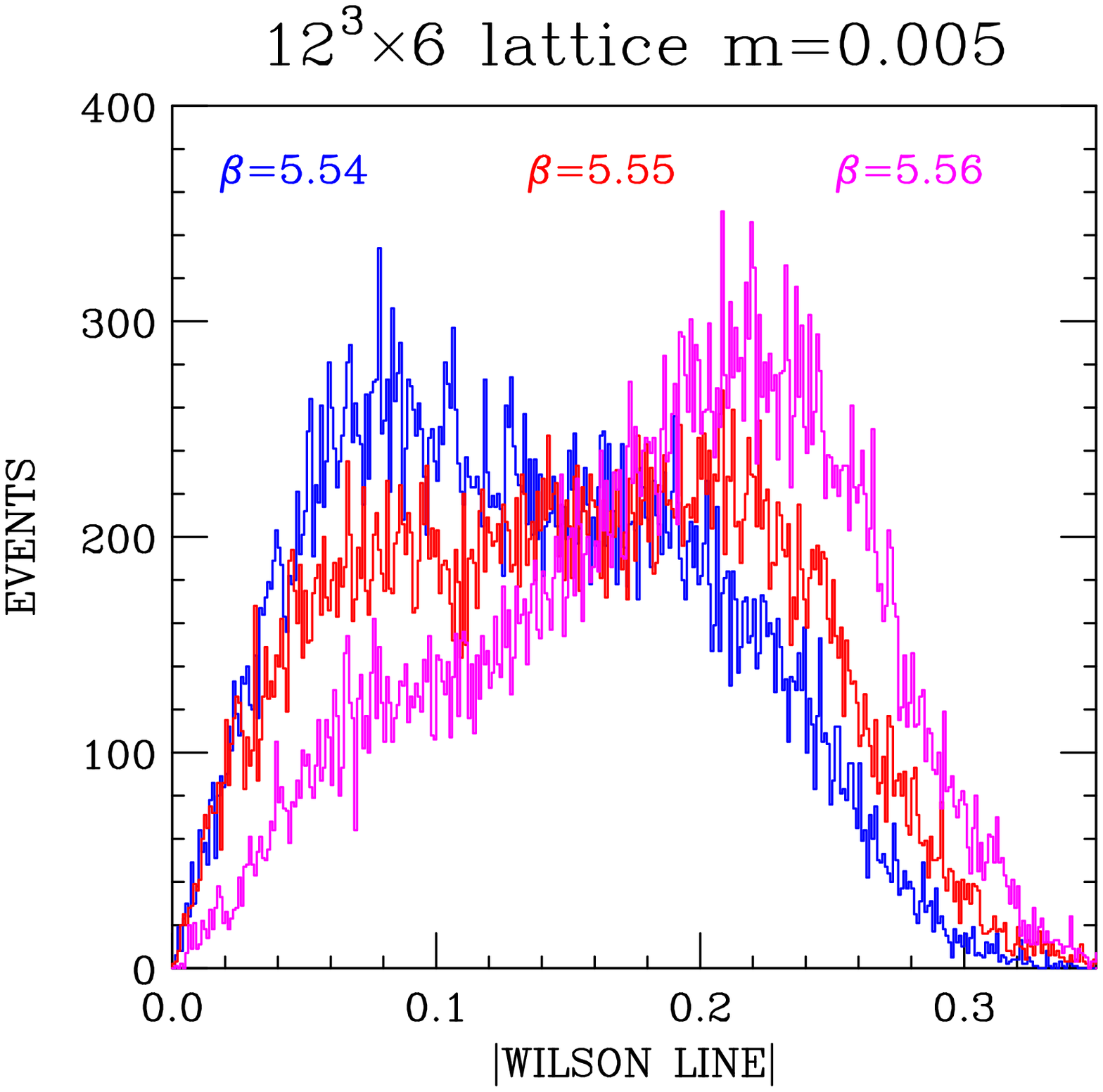}
\epsfxsize=3.2in
\epsffile{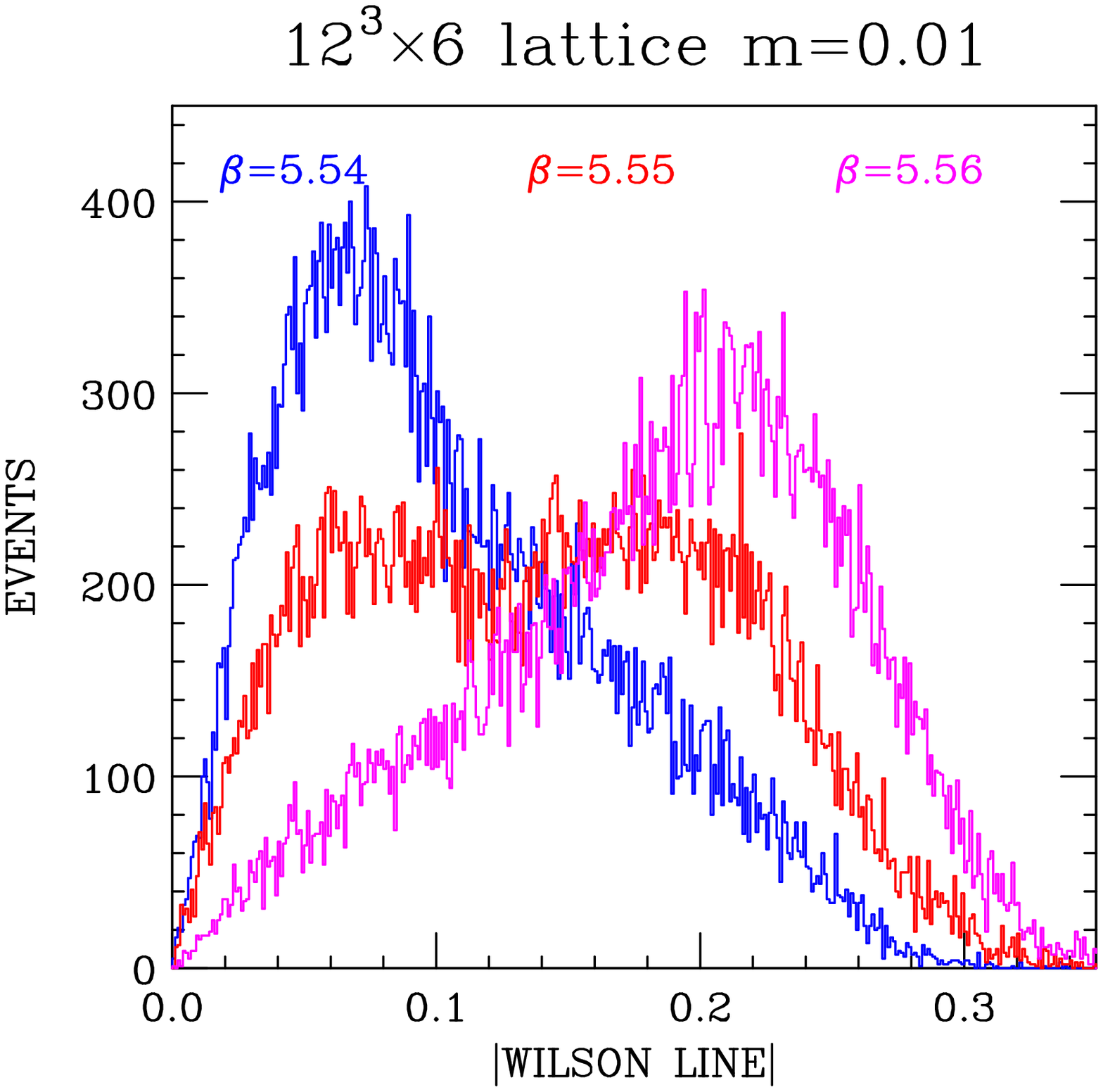}
\epsfxsize=3.2in
\epsffile{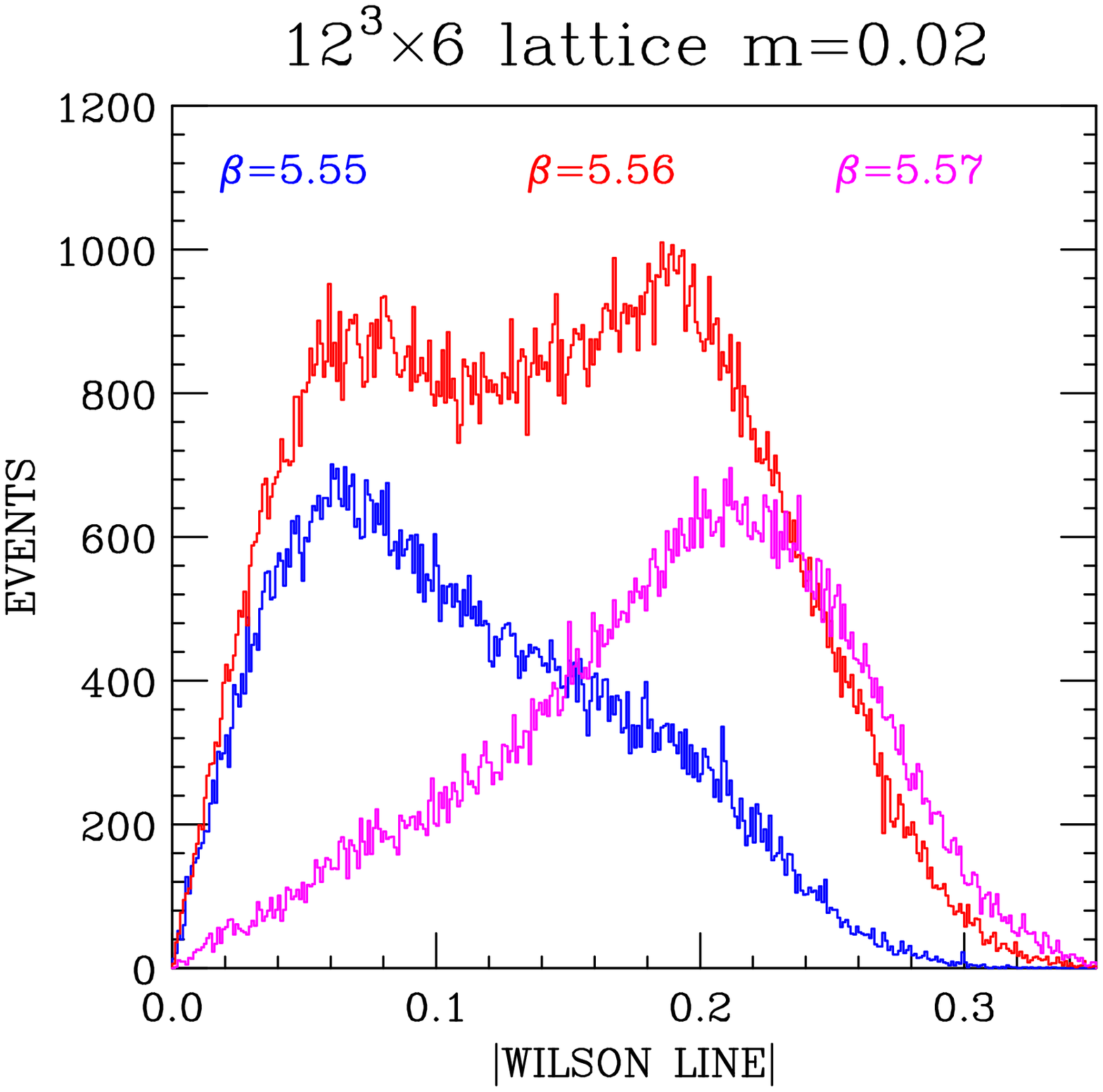}
\caption{Histograms of the magnitudes of Wilson Lines for $\beta$ values 
bracketing the deconfinement transition on a $12^3 \times 6$ lattice for 
a) $m=0.005$, b) $m=0.01$, c) $m=0.02$.}
\label{fig:wilhist}
\end{figure}
To determine the position of the deconfinement transition we examine histograms
of the magnitudes of the Wilson Line close to the transition for each mass.
Although the explicit breaking of the $Z_3$ centre symmetry means that the
magnitudes of the positive and complex Wilson Lines are not identical, they
are close enough in the vicinity of the deconfinement transition that this
fact can be ignored. Such histograms are presented in figure~\ref{fig:wilhist}.
For the lower 2 masses the histogram for each beta represents 50,000 
trajectories. At $m=0.02$ the histograms for $\beta=5.55$ and $\beta=5.57$
represent 100,000 trajectories each, while that at $\beta=5.56$ is for 200,000
trajectories. In each case the histogram peaks at a low value below the
transition and an appreciably higher value above the transition. Very close to
the transition the peak of the histogram is relatively flat, with some hint of
a double peaked structure. From these graphs we estimate that the transition
$\beta$s ($\beta_d$) are $\beta_d(m=0.005)=5.545(5)$,
$\beta_d(m=0.01)=5.550(5)$ and $\beta_d(m=0.02)=5.560(5)$, respectively. As in
the $N_t=4$ case, the mass dependence is weak.

\begin{figure}[htb]
\epsfxsize=5.0in
\epsffile{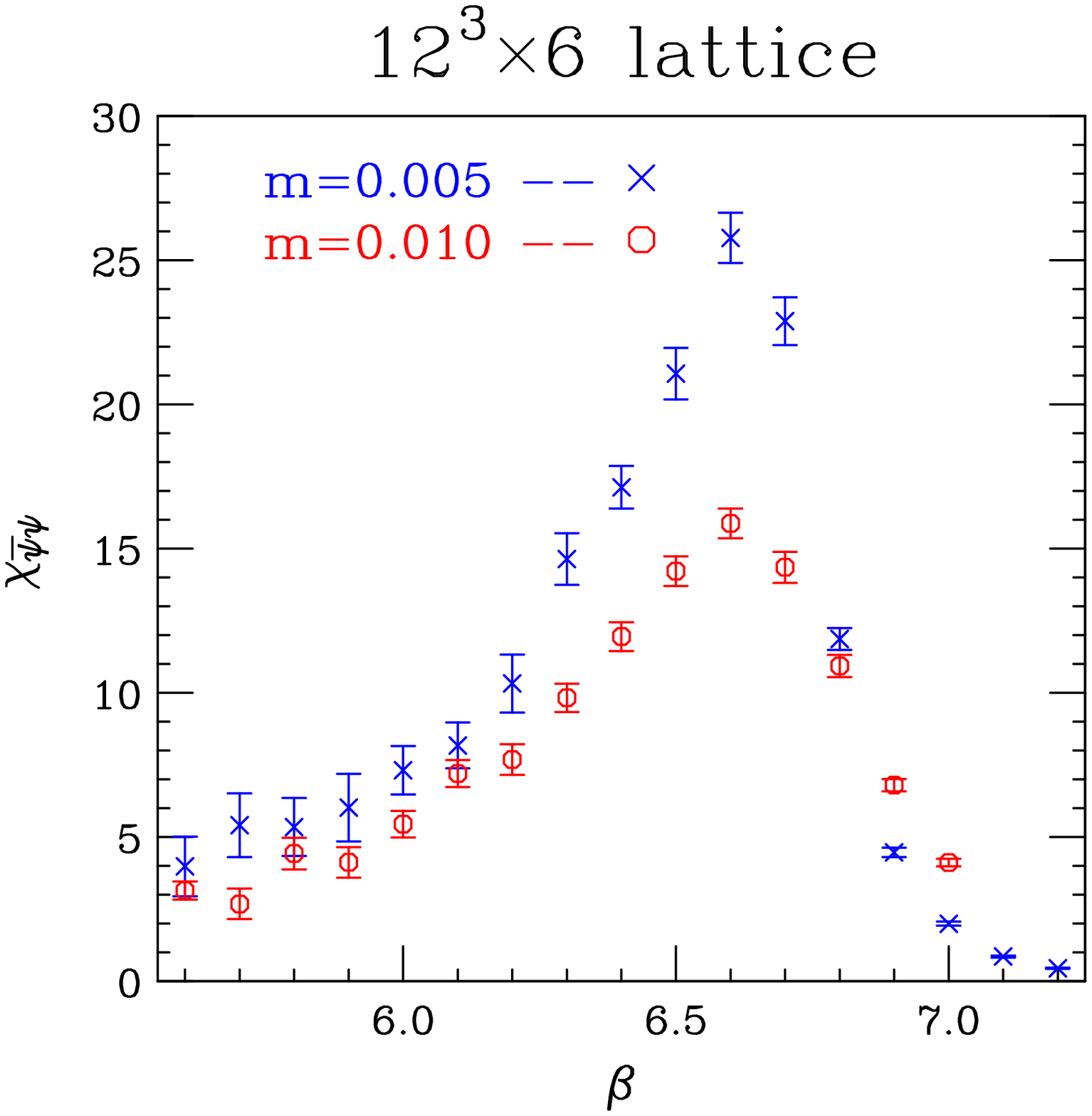}                                                       
\caption{Chiral susceptibilities $\chi_{\bar{\psi}\psi}$ as functions of $\beta$
on a $12^3 \times 6$ lattice for $m=0.005,0.01$, for a $\beta$ range which     
includes the chiral transition.}                                                
\label{fig:chi6}                                                               
\end{figure}
Again we estimate the position of the chiral-symmetry restoration transition by
examining the peaks in the chiral susceptabilities for $m=0.005,0.01$. As for
$N_t=4$ we obtain estimates of the chiral condensate from 5 independent noise
vectors, which yields an unbiased estimate for the chiral susceptibility. These
chiral susceptibilities are plotted versus $\beta$ in figure~\ref{fig:chi6} for
each of these lowest two masses. Again, the $\beta$s we use in this region are
two sparse to allow the use of Ferrenberg-Swendsen reweighting to interpolate
between them. Since our estimate for the peak of the susceptibility plots for
each mass is $\beta=6.60(5)$, we estimate for the position of the chiral
transition at $m=0$ is $\beta=\beta_\chi=6.6(1)$. Note that this estimate is
for the states with positive real Wilson Lines only. 

\begin{figure}[htb]
\epsffile{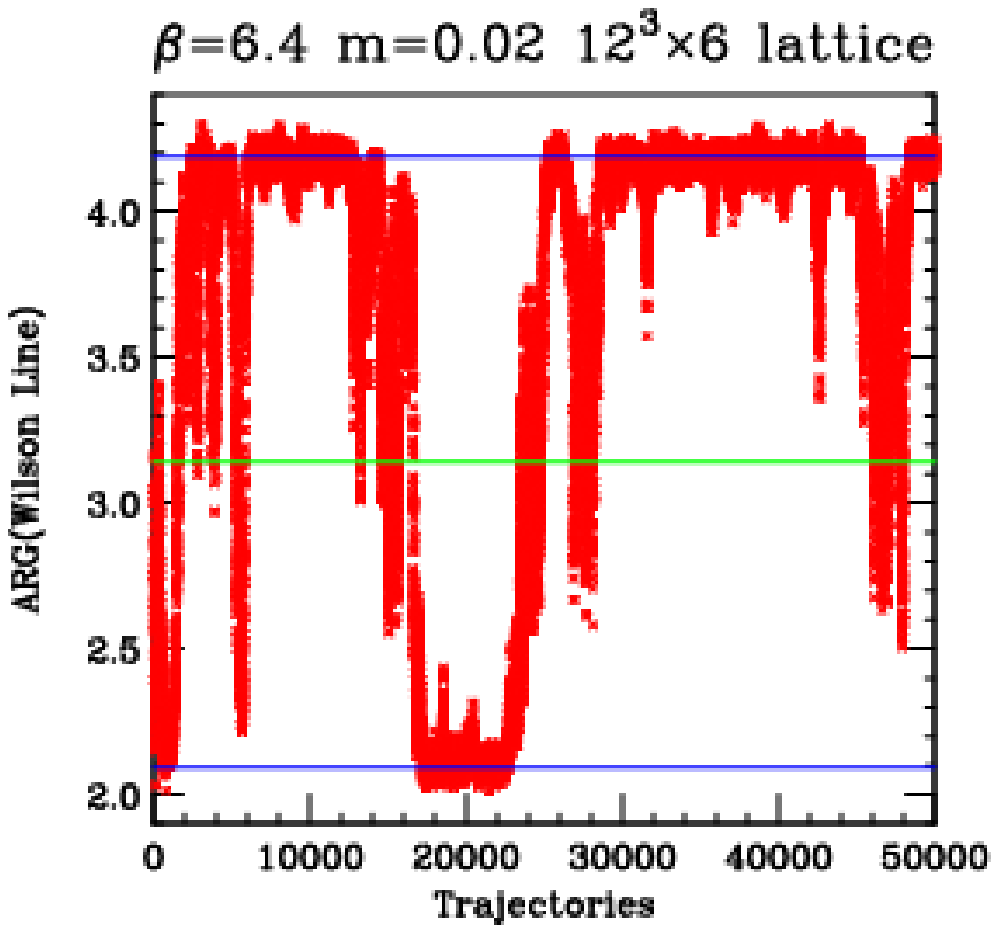}
\epsffile{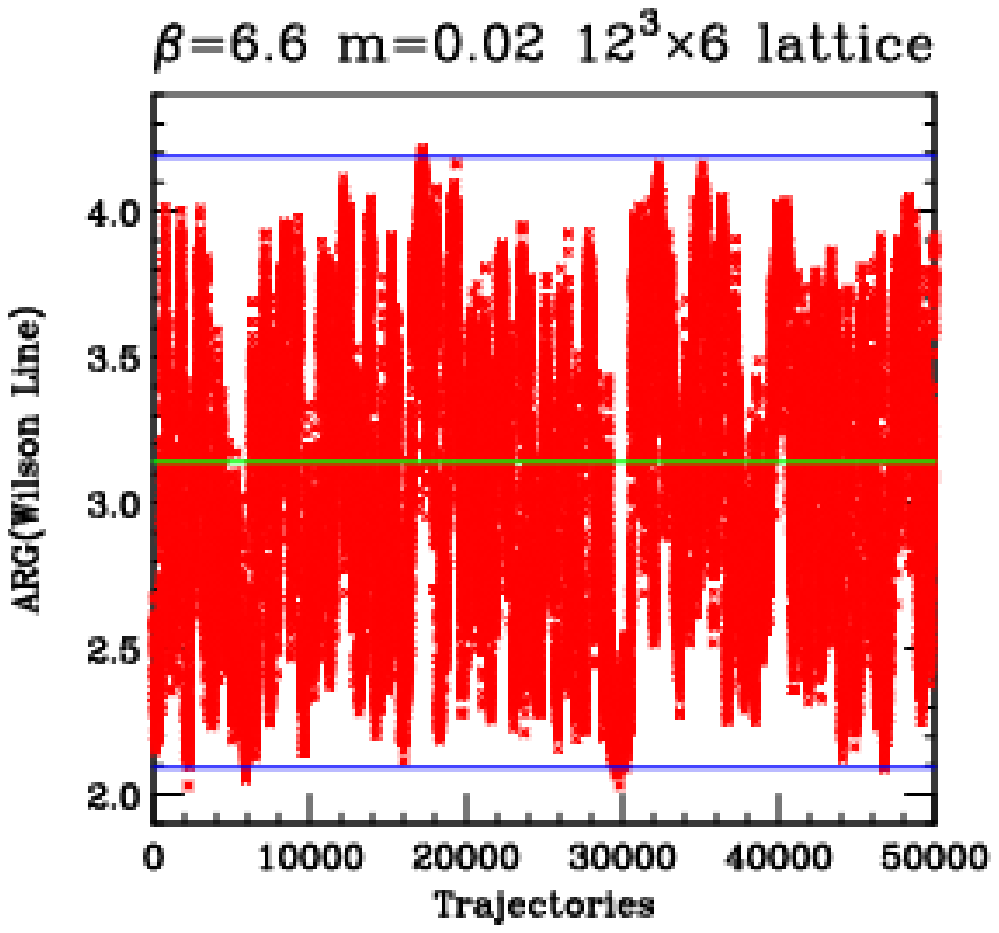}
\caption{The `time' evolution of the argument(phase) of the Wilson Line for 
$m=0.02$ on a $12^3 \times 6$ lattice, a) for $\beta=6.4$ and b) for 
$\beta=6.6$. The horizontal lines are at $2\pi/3$, $\pi$ and $4\pi/3$.}
\label{fig:neg.to.complex}
\end{figure}
Starting at $\beta=7.0$, $m=0.02$ with a real negative Wilson Line, we find
that this state is stable for runs of up to 50,000 trajectories for $\beta$s 
down to $\beta=6.6$. Decreasing this to $\beta=6.4$, we find that the system 
has transitioned to a state with a Wilson Line oriented in the direction of one
of the complex cube roots of unity. Figure~\ref{fig:neg.to.complex} shows
the time evolution of the phases of the Wilson Lines at $\beta=6.4$ and 
$\beta=6.6$ for $m=0.02$. We see that the way the transition proceeds is that 
the fluctuations in the phase become large as $\beta$ approaches the transition
from above until eventually they reach $2\pi/3$ and $4\pi/3$($-2\pi/3$). When
this occurs the system spends appreciably more time at these two values that at
intermediate values. We can then consider the system as being in a state with 
its Wilson Line oriented in the direction of one of the two complex cube roots
of unity. On the finite lattice it tunnels between these 2 states. Approaching
this transition from below we could describe it as the disordering of the 2
states with Wilson Lines in the directions of the complex cube roots of unity,
as suggested by Machtey and Svetitsky. The position of this transition for
$m=0.02$ is at $\beta \approx 6.5$, while for $m=0.01$ it occurs at $\beta
\approx 6.4$.

\section{Discussion and conclusions}

We simulated QCD with two flavours of staggered quarks at finite temperatures
using the RHMC method. Our simulations were performed on $8^3 \times 4$,
$12^3 \times 4$ and $12^3 \times 6$ lattices with 3 quark masses, to allow
for chiral extrapolations. We find widely separated deconfinement and
chiral-symmetry restoration transitions. Both the deconfinement and chiral
transitions move to significantly lower couplings as $N_t$ is increased from
4 to 6, which is the expected behaviour for finite temperature transitions in
an asymptotically free theory. This suggests that the theory is confining
with spontaneously-broken chiral symmetry, while being under the control of
the weak-coupling asymptotically-free ultraviolet fixed point, i.e. that it 
walks.

The deconfinement transition occurs at $\beta=\beta_d$ where for $N_t=4$,
$\beta_d(m=0.02)=5.420(5)$, $\beta_d(m=0.01)=5.4115(5)$ and
$\beta_d(m=0.005)=5.405(5)$ while for $N_t=6$, $\beta_d(m=0.02)=5.560(5)$,
$\beta_d(m=0.01)=5.550(5)$ and $\beta_d(m=0.005)=5.545(5)$. The chiral-symmetry
restoration transition occurs at $\beta=\beta_\chi$, where $\beta_\chi=6.3(1)$
at $N_t=4$ and $\beta_\chi=6.6(1)$ at $N_t=6$.

The large separation of the deconfinement and chiral transitions indicates that
the enhanced attraction between quark-antiquark pairs over that for fundamental
quarks due to the larger quadratic Casimir operator for the sextet
representation ($10/3$ versus $4/3$), causes a chiral condensate to form at a
distance much shorter than the scale of confinement. This effectively removes
the quarks from consideration at longer distances where the theory will behave
more like a pure glue (quenched) theory. This is presumably why, in the
deconfined phase, we see a three state system, where the Wilson Lines align in
the directions of one of the cube roots of unity, a relic of the now-broken
$Z_3$ symmetry. The breaking of $Z_3$ is seen in the difference in magnitudes
of the real positive Wilson Lines versus those with phases close to $\pm
2\pi/3$. At $N_t=4$, the states with complex Wilson Lines are only metastable,
while at $N_t=6$ all 3 states appear to be stable. The existence of states
with all 3 $Z_3$ Wilson Line orientations has been predicted by Machtey and
Svetitsky \cite{Machtey:2009wu} who observed them in their simulations with 2
flavours of colour-sextet Wilson quarks. They also observed the metastability
of the states with complex Wilson Lines. Earlier simulations by DeGrand,
Shamir and Svetitsky had reported deconfined states with Wilson Lines oriented
in the directions of the complex cube roots of unity \cite{DeGrand:2008kx}.

More work needs to be done to determine whether the chiral-symmetry restoring
transition of these complex/negative Wilson Line states is coincident with
that of the state with a positive Wilson Line. In addition we would like to
know whether this chiral transition shows the scaling properties of the
expected $O(2)/O(4)$ universality class. The fact that this transition occurs
at relatively weak coupling should help us in this endeavour.
 
We have also observed an additional transition. At $\beta \approx 5.9$ on an
$N_t=4$ lattice with $m=0.02$ or $m=0.01$, the states with complex Wilson Lines
disorder to a state with a negative Wilson Line. Such a transition is also
observed for $N_t=6$ at $\beta \approx 6.5$ for $m=0.02$ and 
$\beta \approx 6.4$ for $m=0.01$. The existence of states with negative Wilson
Lines and of such a transition is predicted and observed by Machtey and
Svetitsky. Such a transition would be expected to be either a second-order
phase transition in the universality class of the 3-dimensional Ising model,
or a first-order phase transition. The large increase in the $\beta$ for this
transition between $N_t=4$ and $N_t=6$ makes us suspect that this is a lattice
artifact. We also note that by $N_t=6$, it is close to the chiral transition
and could well merge with it at larger $N_t$. Since the negative Wilson Line
state (phase $\pi$) comes from disordering the two states with phases $\pm
2\pi/3$ (a fact also predicted by Machtey and Svetitsky), the magnitude of the
Wilson Line above the transition is approximately half what it is below the
transition. Just below the transition, the magnitude of Wilson Line in the
complex Wilson Line states is approximately $2/3$ that of the positive Wilson
Line state, so that after the transition the magnitude of the negative Wilson
Line is approximately $1/3$ of that of the positive Wilson Line. This would
suggest that the transition might be associated with the symmetry breaking
$SU(3) \rightarrow SU(2) \times U(1)$.

We need to understand why we see well-separated deconfinement and chiral
transitions with staggered fermions, whereas DeGrand, Shamir and Svetitsky
observed these transitions to be coincident with Wilson fermions. Of course, 
since we are far from the weak-coupling limit, different fermion actions do
not have to have the same physics. This is especially true, if we happen to be
in the strong-coupling domain, beyond the infrared fixed point of a conformal
field theory. Of course, the observation from our simulations that the coupling
at each of the transitions is decreasing as the lattice spacing is decreasing
would appear to exclude this possibility, since the $\beta$ function becomes
positive above this fixed point. However, it has recently been suggested that
there could be two non-trivial fixed points \cite{Kaplan:2009kr} in such
theories. In this case, if we are beyond the second non-trivial fixed point
(which would be an ultraviolet fixed point), the coupling would decrease at
short distances. Of course, if we are beyond the region where the ultraviolet
behaviour of the theory is controlled by asymptotic freedom, drawing any
conclusions is pure speculation.

To better understand our results, and to help determine whether the theory is
indeed walking, rather than conformal as indicated by the work of DeGrand,
Shamir and Svetitsky using Wilson fermions, we are now extending our
simulations to $N_t=8$, where finite lattice spacing effects should be
reduced, to see if the deconfinement and chiral-symmetry restoration 
transitions remain consistent with being finite-temperature transitions.
We also plan to perform simulations at zero temperature, measuring
the chiral condensate, the hadron spectrum, $f_\pi$, etc., to test other
aspects of the theory which should help us determine whether this theory is
conformal or walking. In addition we will try to determine the running of a
suitably-defined renormalized coupling. Having two different spatial lattice
sizes at $N_t=4$ showed us that finite size effects are small. We need a
second spatial lattice size for $N_t=6$. We have learned that the authors of
reference \cite{Fodor:2008hm} are now starting simulations of this theory
using improved staggered fermions, which should help resolve these issues
\cite{KN}.

      We are now performing simulations of lattice QCD with 3 flavours of
staggered quarks at finite temperature. Since this theory is almost certainly
conformal, it is interesting to determine whether its behaviour is
qualitatively different from that of the $N_f=2$ case, and whether our
simulations can see this conformality.

\section*{Acknowledgements}

DKS is supported in part by the U.S. Department of Energy, Division of High
Energy Physics, Contract DE-AC02-06CH11357, and in part by the
Argonne/University of Chicago Joint Theory Institute. JBK is supported in part
by NSF grant NSF PHY03-04252. These simulations were performed on the Cray XT4,
Franklin at NERSC under an ERCAP allocation, and on the Cray XT5, Kraken at
NICS under an LRAC/TRAC allocation.

DKS thanks J.~Kuti, D.~Nogradi and F.~Sannino for helpful discussions.

\end{document}